\documentclass[aps,prb,twocolumn]{revtex4-1}

\usepackage[english]{babel}
\usepackage[utf8]{inputenc}
\usepackage{amsmath}
\usepackage{amssymb}
\usepackage[caption=false]{subfig}
\usepackage{amssymb}
\usepackage{epsfig}
\usepackage{graphicx}
\usepackage{amsmath}
\usepackage{array,color}
\usepackage{natbib}

\usepackage{soul}

\usepackage[usenames,dvipsnames]{xcolor}
\definecolor{forestgreen}{rgb}{0.11,0.54,0.15}
\definecolor{purple}{rgb}{0.62,0.10,0.96}
\definecolor{dockerblue}{rgb}{0.11,0.56,0.98}
\definecolor{freeblue}{rgb}{0.25,0.41,0.88}

\usepackage[pdftex,plainpages=false,colorlinks=true,linkcolor=Red, citecolor=blue, urlcolor=blue]{hyperref}

\begin{document}

\title{Mott transition and high-temperature crossovers at half-filling}

\author{A. Reymbaut$^{1,2}$, M. Boulay$^1$, L. Fratino$^{3,4}$, P. Sémon$^{1,5}$, Wei Wu$^1$, G. Sordi$^3$, and A.-M. S. Tremblay$^{1,2}$}
\affiliation{
$^1$Département de physique and Regroupement québécois sur les matériaux de pointe,
Université de Sherbrooke, Sherbrooke, Québec, Canada J1K 2R1 \\
$^2$Institut quantique, Université de Sherbrooke, Sherbrooke, Québec, Canada J1K 2R1 \\
$^3$Department of Physics, Royal Holloway, University of London, Egham, Surrey TW20 0EX, United Kingdom \\
$^4$ Laboratoire de Physique des Solides, CNRS UMR 8502, Univ. Paris-Sud, Université Paris-Saclay F-91405 Orsay Cedex, France \\
$^5$Computational Science Initiative, Brookhaven National Laboratory, Upton, New York 11973-5000, USA\\
}


\date{\today}

\begin{abstract}
The interaction-driven Mott transition in the half-filled Hubbard model is a first-order phase transition that terminates at a critical point $(T_\mathrm{c},U_\mathrm{c})$ in the temperature-interaction plane $T-U$. A number of crossovers occur along lines that extend for some range above $(T_\mathrm{c},U_\mathrm{c})$. Asymptotically close to $(T_\mathrm{c},U_\mathrm{c})$, these lines coalesce into the so-called Widom line. The existence of $(T_\mathrm{c},U_\mathrm{c})$ and of the associated crossovers becomes unclear when long-wavelength fluctuations or long-range order occur above $(T_\mathrm{c},U_\mathrm{c})$. 
We study this problem using continuous-time quantum Monte Carlo methods as impurity solvers for both Dynamical Mean-Field Theory (DMFT) and Cellular Dynamical Mean-Field Theory (CDMFT).
We contrast the cases of the square lattice, where antiferromagnetic fluctuations dominate in the vicinity of the Mott transition, and the triangular lattice where they do not. The inflexion points and maxima found near the Widom line for the square lattice can serve as proxy for the triangular lattice case. But the only crossover observable in all cases at sufficiently high temperature is that associated with the opening of the Mott gap. The same physics also controls an analog crossover in the resistivity called the ``Quantum Widom line".
\end{abstract} 


\email[email:]{andre-marie.tremblay@usherbrooke.ca}

\maketitle



\section{Introduction}

In certain materials, generally those containing $d$ or $f$ electrons, bandwidths can become smaller than electronic repulsion, leading to an interaction-driven metal-insulator transition in a half-filled band: the Mott transition. A Mott-insulating state is typically observed at high temperature in materials that band theory suggests should be metallic. A case in point is the high-temperature insulating state of cuprate superconductors that exists above an antiferromagnetic phase at half-filling.\cite{anderson1987resonating} At temperatures that are low, but not low enough for a given system to exhibit long-range order, one can sometimes observe a Mott transition by increasing the interaction strength to bandwidth ratio. This occurs as a first-order transition in, for example, layered organic superconductors when one decreases the pressure,\cite{Lefebvre:2000,Limelette:2003,Powell:2006b,FurukawaWidom:2015,Pustogow:2018}  which decreases the bandwidth.\footnote{The possible role of phonons in this context has been discussed in Ref.~\onlinecite{hassan2005sound}}

The Mott transition is without any doubt an important feature of strongly correlated materials, but the high-temperature regime above this celebrated transition is as interesting, since there the phase diagram becomes more universal. Indeed, systems as diverse as three-dimensional transition metal oxide alloys and two-dimensional organic materials exhibit the same qualitative high-temperature phase diagram.\cite{Boer:1937,Rice:1973,Powell:2005}

Dynamical Mean-Field Theory (DMFT)~\cite{Georges:1996} has provided a framework to understand the Mott transition. In the original approach, a single site is embedded in a self-consistent bath of non-interacting electrons that provides a dynamical mean field that leads to a first-order metal-insulator transition. This demonstrates that the Mott transition has nothing to do with long-range electronic correlations. In the case where there are long-wavelength fluctuations, the Mott transition may be hidden, as shown by studies of the Hubbard model on the square lattice that used a number of advanced methods.~\cite{Aichhorn:2014} Nonetheless, it has been understood from the start~\cite{Georges:1996} that the Mott transition is meant to describe the physics at high temperature, above the regime of long-wavelength fluctuations of collective modes. 

In the high-temperature regime of correlated materials, one observes several crossovers that can be understood as consequences of a Mott transition that may not occur in the true ground state of the system. When the ground state does not feature a Mott transition, this transition and its second-order critical end point can still be understood as unphysical fixed points that control the physical high-temperature regime. 
The main crossovers characterizing this regime can be summarized as follows. In the supercritical regime above the critical end point of the Mott transition, there is a line terminating at the critical end point, along which various crossovers merge: the so-called \textit{Widom line}. While this terminology was introduced in strongly correlated electron physics in Ref.~\onlinecite{Sordi:2012}, it was originally defined for classical liquid-gas transitions in Ref.~\onlinecite{XuStanleyWidom:2005}. There is another crossover that occurs along what we call the \textit{Mott line}, that determines at which temperature an insulating Mott gap is fully formed. A crossover closely related to the Mott line, observed in the resistivity of strongly correlated systems, has been dubbed the \textit{Quantum Widom line}.\cite{Dobrosavljevic:2013,Vucicevic_Tanaskovic:2015,Eisenlohr_Lee_Vojta_2019}

Here, we study the single-band Hubbard model with nearest-neighbor hopping on two lattices, the square and the isotropic triangular lattices at half-filling. In addition to DMFT, we use the Cellular Dynamical Mean-Field Theory (CDMFT).\cite{Maier:2005} Our contribution~\cite{reymbaut_phd_thesis} is to show which features of the Mott transition are observable at high temperature when the system has strong antiferromagnetic fluctuations, as on the square lattice, or when frustration allows the Mott transition to manifest itself directly, as on the triangular lattice. We also perform some benchmarking across various numerical techniques to identify the temperature regime within which these techniques describe the same physics. The Mott line is the only feature that is universally observable at high temperature, independently of frustration.\cite{reymbaut_phd_thesis} 

Note that in the case of the square lattice at small interaction strength, a number of crossovers have been identified as temperature is decreased towards zero.~\cite{Kim_Simkovic_Kozik_2020} We focus on the crossovers associated with the Mott transition, not on those crossovers.

In the following section, we describe the model and the methods. 
Section~\ref{Sec_Crossovers_characterization} introduces definitions of the various crossover lines of interest that do not require analytic continuation. Our results are discussed in Section~\ref{Sec_Discussion} and in the conclusion. Section~\ref{Sec_Discussion} includes analytically continued results that provide a deeper understanding of the Widom and Mott lines while also justifying the definitions of Section~\ref{Sec_Crossovers_characterization}. 

\section{Model and methods}
\label{Sec_model_methods}

The physics of the Mott transition is controlled by the competition between kinetic energy, represented by nearest-neighbor hopping $t\equiv 1$ on a lattice, and purely local screened Coulomb repulsion $U$. This physics is contained in the two-dimensional single-band Hubbard model:\cite{Gutzwiller:1963,Hubbard:1963,Kanamori:1963} 
\begin{equation}
\hat{\mathcal{H}} = -t\sum_{\langle i, j\rangle \; \sigma} \left(\hat{c}^\dagger_{i\sigma} \, \hat{c}_{j\sigma}+ \mathrm{h.c}\right) + U \sum_{i} \hat{n}_{i\uparrow} \, \hat{n}_{i\downarrow} - \mu \sum_i \hat{n}_i,
\end{equation}
where $\hat{c}^{(\dagger)}_{i\sigma}$ is the fermionic destruction (creation) operator at site $i$ and spin $\sigma$, $\hat{n}_{i\sigma} = \hat{c}^{\dagger}_{i\sigma}\hat{c}_{i\sigma}$ is the number operator (with $\hat{n}_i = \hat{n}_{i\uparrow} + \hat{n}_{i\downarrow}$), and $\mu$ is the chemical potential.

Although simple mean-field theory may perform well when interactions are weak and temperature is high, the study of stronger interactions requires the use of improved versions of this theory in order to uncover the physics on a larger energy scale. These approaches include Dynamical Mean-Field Theory (DMFT),\cite{Georges:1996} where a single site is dynamically coupled to a bath of non-interacting electrons, and Cellular Dynamical Mean-Field Theory (CDMFT),\cite{Maier:2005,KotliarRMP:2006,LTP:2006} where it is a cluster of sites that is coupled to a non-interacting bath. It is then the frequency dependence of the hybridization between cluster and bath that becomes essential. The solution of the resulting quantum-impurity problems is usually obtained with an ``impurity solver''. While exact diagonalization is mostly used at zero temperature, continuous-time quantum Monte Carlo algorithms (CTQMC)~\cite{Gull:2011} are preferred at finite temperature. Here, we use the interaction-expansion algorithm (CT-INT), also known as the Rubtsov algorithm, especially suited when interactions are weak to intermediate,\cite{Rubtsov:2005} and the hybridization-expansion algorithm (CT-HYB),\cite{werner:2006,WernerMillis:2006,haule:2007i} especially suited when interactions are strong.\cite{Gull:2007} We also use the Dynamical Cluster Approximation (DCA)\cite{Maier:2005} and the Determinant Quantum Monte Carlo (DQMC)\cite{Blankenbecler:1981,LOH-DQMC:1992} for benchmarking purposes.

We study various lattice geometries to highlight the effects of cluster size and magnetic frustration on the high-temperature crossovers. We take the square lattice as an example of a lattice that is not magnetically frustrated and we study it with a four-site ($2\times 2$) square cluster using CDMFT (CDMFT $2\times 2$). We take the triangular lattice as an example of a fully frustrated lattice and we study it using a three-site isotropic triangular cluster with CDMFT (CDMFT Triangle).
%
%
It is also useful to compute the results of single-site DMFT. Indeed, the DMFT results should be close to those of CDMFT Triangle since the local nature of DMFT implicitly assumes that the lattice is magnetically frustrated so that long-wavelength magnetic correlations are unimportant. In general, one expects long-wavelength magnetic correlations to be present at lower temperatures for the frustrated lattice than for the non-frustrated one. In three dimensions, long-range order would occur at lower temperature for a frustrated system. This is important since this is what leads to the possibility of observing the Mott transition experimentally. 

Table~\ref{Table_numerical_techniques} sums up the various methods and their associated abbreviations in this work. 

\begin{widetext}

\begin{table}[h]
\begin{tabular}{|c|c|c|}
\hline
\textbf{Chosen abbreviation} & \textbf{Numerical technique} & \textbf{Impurity solver} \\
\hline
\hline
DMFT & DMFT using a square lattice dispersion & CT-INT (at low interaction) \\
 &  & CT-HYB (at high interaction) \\
\hline
CDMFT $2\times 2$ & CDMFT on a four-site square cluster & CT-HYB \\
\hline
CDMFT Triangle & CDMFT on a three-site isotropic triangular cluster & CT-HYB \\
\hline
\end{tabular}
\begin{center}
DMFT: Dynamical Mean-Field Theory\\
CDMFT: Cellular Dynamical-Mean Field Theory\\
CT-HYB: continuous-time quantum Monte Carlo with hybridization expansion\\
CT-INT: continuous-time quantum Monte Carlo with interaction expansion (Rubtsov algorithm) 
\end{center}
\caption{Summary of the different dynamical mean-field theories and impurity solvers used in this work.}
\label{Table_numerical_techniques}
\end{table}


\end{widetext}

\section{Definitions of the crossover lines}
\label{Sec_Crossovers_characterization}

In the following two subsections, we define in turn different versions of the Widom line, and of the Mott line, introduced in the previous section. 


\subsection{Widom lines} 

The Widom line is defined as an inflexion point in thermodynamic potentials. It is found in supercritical fluids by observing where the lines of maxima for different response functions (derivatives of thermodynamic potentials) asymptotically coalesce into one at the approach of the critical point.\cite{XuStanleyWidom:2005, Stanley:2010} In strongly correlated electronic systems, the double occupancy $D=\langle \hat{n}_{i \uparrow}\, \hat{n}_{i \downarrow}\rangle$, \textit{i.e.} the probability for a given site to be doubly occupied, is a thermodynamic quantity of interest since it measures potential energy. We thus define one of the crossovers associated to the Widom line as a maximum in the derivative of $D$ with respect to $U$, an analog of a response function. For a given temperature, an estimate of the location of this Widom line, $U_\mathrm{W}$, is obtained from
\begin{equation}
U_\mathrm{W} = \mathrm{argmax}_U \left| \frac{\partial D}{\partial U} \right|.
\label{Eq_Widom_D}
\end{equation}

\begin{figure}[h!]
\begin{center}
\includegraphics[width=0.48\textwidth]{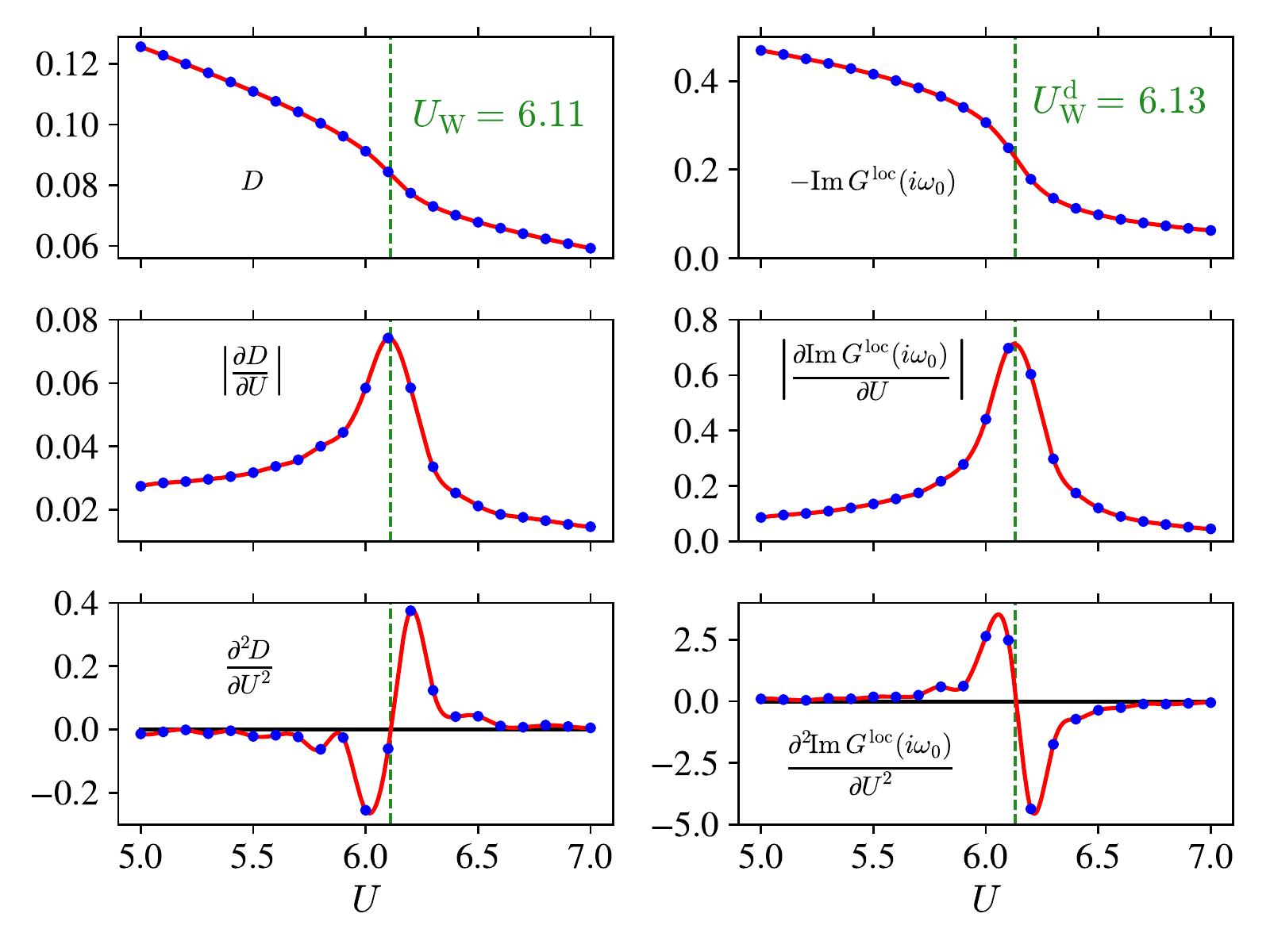}
\caption{\textbf{Left panels :} Determination of the thermodynamical Widom line of CDMFT $2\times 2$ at $\beta=11$ from the behavior of double occupancy. \textbf{Right panels :} Determination of the dynamical Widom line of CDMFT $2\times 2$ at $\beta=11$ from the imaginary part of the Matsubara Green's function at the lowest frequency.}
\label{Fig_Widom_CDMFT_2x2_beta_11}
\end{center}
\end{figure}

We also look for an inflexion point in the imaginary part of the local Green's function $G^\mathrm{loc}$ taken at the first Matsubara frequency $i\omega_0$, which can be thought of as a proxy for the local density of state at $\omega=0$ without requiring analytic continuation. This defines a ``dynamical Widom line" whose location $U_\mathrm{W}^\mathrm{d}$ for a given temperature is found from
\begin{equation}
U_\mathrm{W}^\mathrm{d} = \mathrm{argmax}_U \left| \frac{\partial \, \mathrm{Im}\,G^\mathrm{loc}(i\omega_0)}{\partial U} \right|.
\label{Eq_Widom_G}
\end{equation}
The whole process is illustrated in figure Fig.~\ref{Fig_Widom_CDMFT_2x2_beta_11}.

\subsection{Mott line} 
\label{Sec:Mott_Line}

Finding the opening of a Mott gap while avoiding analytic continuation, which is impossible at high temperature, is a challenge that we address here. When it is not fully opened, a Mott gap differs from a band gap by the the fact that the density of states rearranges itself when the chemical potential varies. There is no rigid band. Nevertheless, when the chemical potential is not too close to the band edge and the gap is large, one can extract the value of the gap by taking advantage of thermal excitations across it, as we proceed to show.  

When the chemical potential lies within a well formed gap, one can assume that the density of states is rigid and we can write down the occupation of a site as
\begin{equation}
n = 2\iint\! \frac{\mathrm{d}^2 \vec{k}}{(2\pi)^2} \int\! \frac{\mathrm{d}\omega}{2\pi} \, \frac{A(\vec{k},\omega)}{e^{\beta(\omega-\mu)}+1} = \int\!\mathrm{d}\omega \, \frac{A^\mathrm{loc}(\omega)}{e^{\beta(\omega-\mu)}+1},
\end{equation}
where the factor 2 comes from spin, $A(\vec{k},\omega)$ is the spectral function, $\beta = 1/T$ is the inverse temperature (with the Boltzmann constant $k_B \equiv 1$), and $A^\mathrm{loc}(\omega)$ is the local density of states
\begin{equation}
A^\mathrm{loc}(\omega) = 2 \iint\! \frac{\mathrm{d}^2 \vec{k}}{(2\pi)^3} \, A(\vec{k},\omega).
\end{equation}
Let us denote by $\mu_-$ and $\mu_+$ the two energies between which a Mott gap opens, with $\mu_- < \mu_+$ so that the gap energy reads $E_\mathrm{g}=\mu_+-\mu_-$. Then the local density of states is zero within the interval $[\mu_-,\mu_+]$ and one has
\begin{equation}
n = \int_{-\infty}^{\mu_-} \! \mathrm{d}\omega\, \frac{A^{loc}(\omega)}{e^{\beta(\omega-\mu)}+1} + \int_{\mu_+}^{+\infty}\! \mathrm{d}\omega\, \frac{A^\mathrm{loc}(\omega)}{e^{\beta(\omega-\mu)}+1}.
\end{equation}
One can then consider two temperature regimes. At zero temperature, the lower Hubbard band is full and the system is half-filled, which gives
\begin{equation}
n = \int_{-\infty}^{\mu_-} \! \mathrm{d}\omega\, A^\mathrm{loc}_{T=0}(\omega) = 1.
\end{equation}
Now, if $T \ll E_\mathrm{g}$, one can approximate
\begin{equation}
\int_{-\infty}^{\mu_-} \! \mathrm{d}\omega\, A^\mathrm{loc}(\omega) \simeq 1
\end{equation}
and write
\begin{eqnarray}
n & \simeq & 1 + \int_{-\infty}^{\mu_-} \! \mathrm{d}\omega\left[\frac{1}{e^{\beta(\omega-\mu)}+1} -1 \right] A^\mathrm{loc}(\omega) \nonumber \\
 & & \quad + \int_{\mu_+}^{+\infty}\! \mathrm{d}\omega\, \frac{A^\mathrm{loc}(\omega)}{e^{\beta(\omega-\mu)}+1},
\end{eqnarray}
or
\begin{equation}
n \simeq 1 + \int_{-\infty}^{\mu_-} \! \mathrm{d}\omega\, \frac{A^\mathrm{loc}(\omega)}{e^{-\beta(\omega-\mu)}+1} + \int_{\mu_+}^{+\infty}\! \mathrm{d}\omega\, \frac{A^\mathrm{loc}(\omega)}{e^{\beta(\omega-\mu)}+1}.
\label{Eq_Mott_line_low_T_limit}
\end{equation}
Recalling that $\omega - \mu < 0$ in the first integral and that $\omega - \mu > 0$ in the second integral, we write
\begin{eqnarray}
n 
& \simeq & 1 + \int_{-\infty}^{0} \! \mathrm{d}\omega\, A^\mathrm{loc}(\omega+\mu_-)\, e^{\beta(\omega + \mu_- -\mu)} \nonumber \\
& & \qquad + \int_{0}^{+\infty}\! \mathrm{d}\omega\, A^\mathrm{loc}(\omega+\mu_+)\, e^{-\beta(\omega + \mu_+ - \mu)}.
\end{eqnarray}
One can finally consider that at low temperature the local density of states varies slowly on the scale of the thermal energy near the band edges and obtain
\begin{eqnarray}
n 
%
& \simeq & 1 + \int_{0}^{+\infty}\! \mathrm{d}\omega\, e^{-\beta\omega} \left[A^\mathrm{loc}(\mu_+)\, e^{-\beta(\mu_+ - \mu)} \right. \nonumber \\
& & \qquad \qquad \qquad \qquad - \left. A^\mathrm{loc}(\mu_-)\, e^{\beta(\mu_- -\mu)} \right]  \\
& \simeq & 1 + T \left[ A^\mathrm{loc}(\mu_+) \,e^{-\beta(\mu_+-\mu)} -  A^\mathrm{loc}(\mu_-) \,e^{\beta(\mu_- -\mu)} \right]. \nonumber
\label{Eq_general_case}
\end{eqnarray}

\begin{figure}[h!]
\begin{center}
\includegraphics[width=0.48\textwidth]{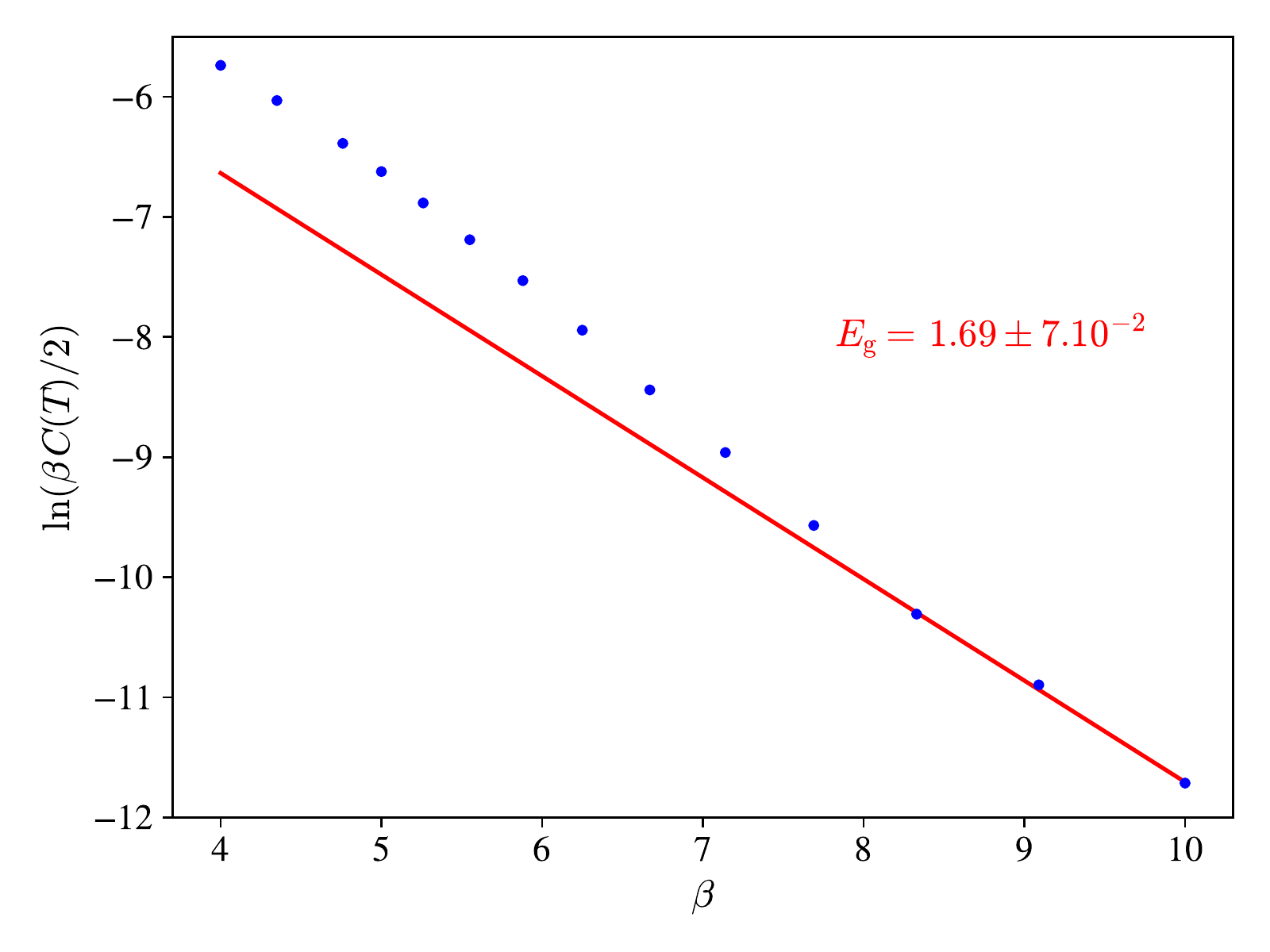}
\caption{Determination of the crossover towards a fully opened Mott gap at $U=8$ in CDMFT $2\times 2$. The fit of Eq.~\eqref{Eq_Fit_gap} seems to hold from $\beta=8.33$ to $\beta=10$, which indicates that the crossover is located at $T \simeq 1/8$ and that the Mott gap must close at a slightly higher temperature. The value of the gap can be estimated from the fit to be $E_\mathrm{g}\simeq 1.69$.}
\label{Fig_Mott_CDMFT_2x2_U_8}
\end{center}
\end{figure}

In the particle-hole symmetric case or on the square lattice with nearest-neighbor hopping, one has $\mu_\pm = \pm E_\mathrm{g}/2$ and the density of states is even in frequency. Thus, one obtains
\begin{eqnarray}
n & \simeq & 1 + 2T\, A^\mathrm{loc}\!\left(\frac{E_{g}}{2}\right)\, e^{-\beta E_{g}/2}\, \sinh\!\left(\beta\mu\right) \nonumber \\
& \simeq & 1 + C(T)\, \sinh\!\left(\beta\mu\right),
\label{Eq_symmetric}
\end{eqnarray}
where 
\begin{equation}
C(T) = 2T\, N\!\left(\frac{E_{g}}{2}\right)\, e^{-\beta E_{g}/2}.
\end{equation}
Fitting Eq.~\eqref{Eq_symmetric} gives $C(T)$. Then, one can fit
\begin{equation}
\ln\left( \frac{\beta \, C(T)}{2} \right) = -\frac{E_\mathrm{g}}{2}\,\beta + \ln \left( A^\mathrm{loc}\!\left(\frac{E_{g}}{2}\right) \right)
\label{Eq_Fit_gap}
\end{equation}
to find the gap energy $E_\mathrm{g}$. This last fit is good only as long as the gap is sufficiently opened for a given temperature so that $T\ll E_\mathrm{g}$, \textit{i.e.} below a temperature lower than the gap-opening temperature. To find the Mott line at a given interaction, one just has to find the temperature above which the fit Eq.~\eqref{Eq_Fit_gap} is no longer good: this means that the gap is about to close. This last point is illustrated in figure Fig.~\ref{Fig_Mott_CDMFT_2x2_U_8}.

\begin{figure}[h!]
\begin{center}
\includegraphics[width=0.48\textwidth]{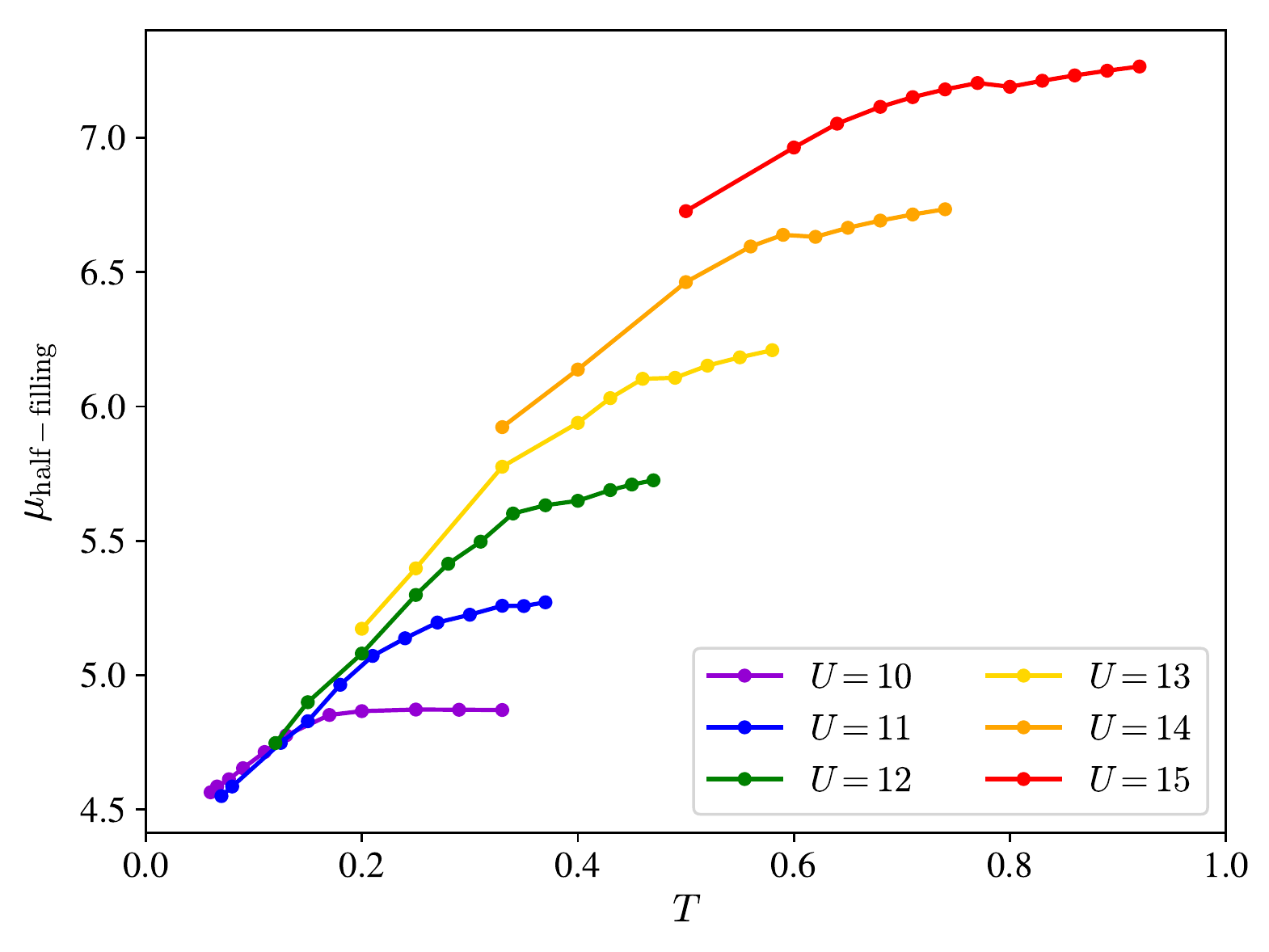}
\caption{Value of the chemical potential at half-filling obtained with CDMFT Triangle for a given interaction $U$ as a function of temperature.
}
\label{Fig_mu_half_filling}
\end{center}
\end{figure}

When there is no particle-hole symmetry, as on a triangular lattice, which is not bipartite, finding the Mott line is not so straightforward. Indeed, one must expect an asymmetric occupation $n$ around the chemical potential at half-filling $\mu_{\mathrm{half-filling}}$, which can also be strongly temperature-dependent, as shown in figure Fig.~\ref{Fig_mu_half_filling}. One has no other choice than to split Eq.~\eqref{Eq_general_case} in two by finding the two situations where one can neglect one exponential compared with the other, in order to isolate the electron-doped ($\mu>\mu_{\mathrm{half-filling}}(T)$) and the hole-doped ($\mu<\mu_{\mathrm{half-filling}}(T)$) cases. The chemical potential cannot be in the middle of the gap because both exponentials of Eq.~\eqref{Eq_general_case} would be comparable, but it cannot be too close to the band edges $\mu_+$ or $\mu_-$ either because the whole band structure would rearrange itself. Keeping that in mind and neglecting one exponential compared to the other in Eq.~\eqref{Eq_general_case}, one obtains
\begin{eqnarray}
\pm(n-1) & \simeq & T\,A^\mathrm{loc}(\mu_\pm) \,e^{\mp\beta\mu_\pm}\,e^{\pm\beta\mu} \nonumber \\
 & \simeq & C_\pm(T)\,e^{\pm\beta\mu},
\end{eqnarray}
where 
\begin{equation}
C_\pm(T) = T\,A^\mathrm{loc}(\mu_\pm) \,e^{\mp\beta\mu_\pm}.
\end{equation}
To find $\mu_+$ and $\mu_-$, and then extract $E_\mathrm{g}=\mu_+-\mu_-$, one can then perform the same fit as that used in the particle-hole symmetric case but for the electron-doped and the hole-doped cases separately. Again, to find the crossover temperature above which the Mott gap starts to close, one just has to find the temperature above which the fits 
\begin{equation}
\ln (\beta\,C_\pm(T)) = \mp\mu_\pm\, \beta + \ln A^\mathrm{loc}(\mu_\pm)
\label{Eq_Fit_Mott_non-symmetric}
\end{equation}
are no longer good. Note that this time, there will in general be two different crossovers, one for each kind of doping.

\section{Results}
\label{Sec_Discussion}

We begin with the results for the frustrated triangular lattice where the Mott transition and its crossovers are least affected by magnetic correlations. The next subsection considers the opposite limit where antiferromagnetic fluctuations dominate in the region where the Mott transition would occur. This allows us to later identify the features of the high-temperature phase diagram that are robust against antiferromagnetic fluctuations. Finally, we show results obtained through analytic continuation that help understand the meaning of the Widom and Mott lines, and justify the criteria detailed in Section~\ref{Sec_Crossovers_characterization}. 

\subsection{Triangular lattice} 

\begin{figure}[h!]
\begin{center}
\includegraphics[width=0.48\textwidth]{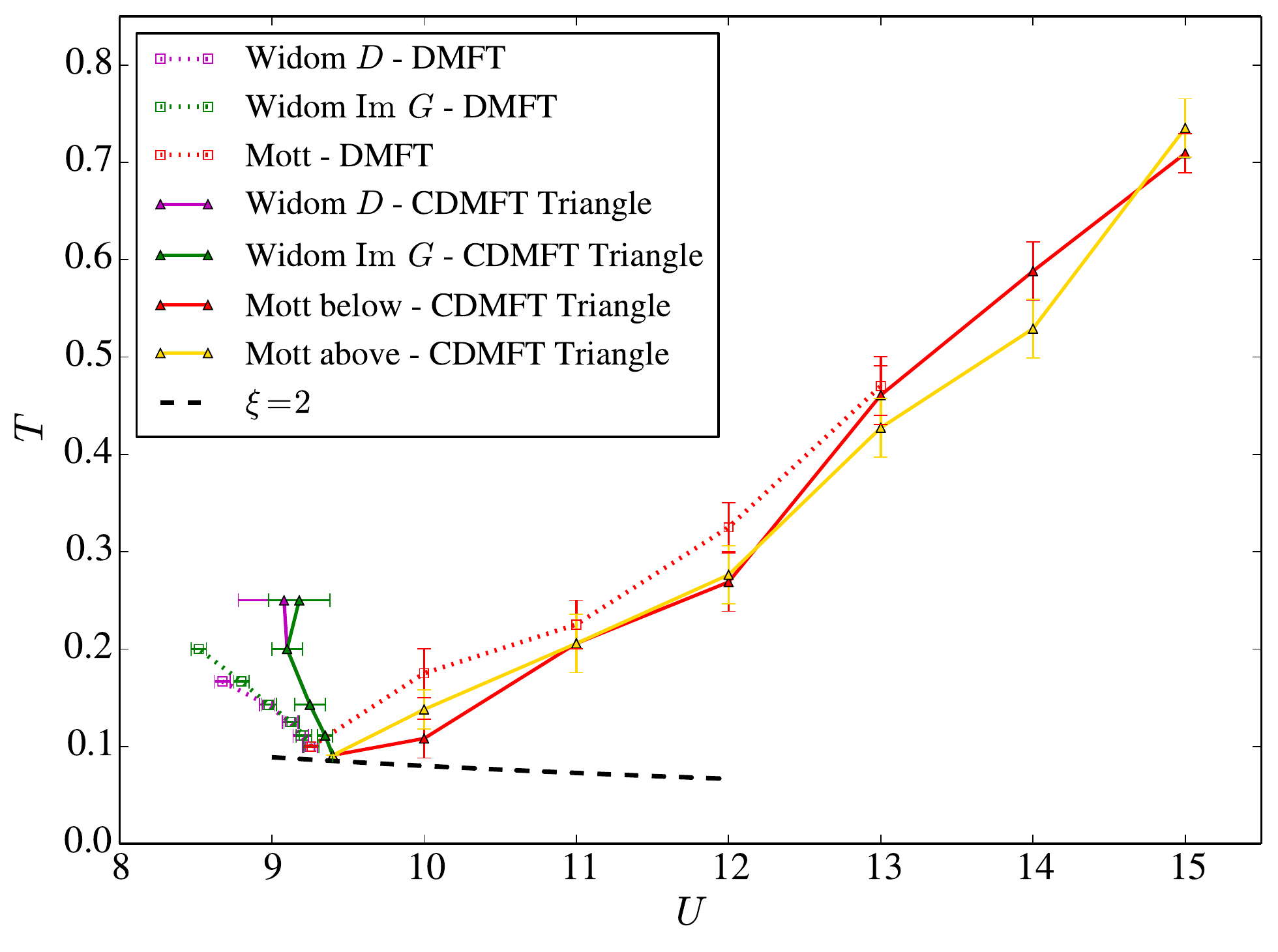}
\caption{$T-U$ phase diagram for DMFT and CDMFT Triangle. ``Widom $D$'' and ``Widom $\mathrm{Im}\,G$'', correspond to the Widom lines associated to an inflexion point in $D$ and in $\mathrm{Im}\,G^\mathrm{loc}(i\omega_0)$, respectively. A recent dual fermion calculation finds an inflection point in $D$ at $T=1/6$ that is consistent with CDMFT Triangle.~\cite{Li_Gull_2020} The other lines, dubbed ``Mott'', stand for the Mott lines associated to the opening of the Mott gap. There are two such crossovers for CDMFT Triangle: one found with values of $\mu$ ``above" $\mu_{\mathrm{half-filling}}(T)$ and one found with values of $\mu$ ``below" $\mu_{\mathrm{half-filling}}(T)$, in accordance with Eq.~\eqref{Eq_Fit_Mott_non-symmetric}. The dashed line indicates where the magnetic correlation length $\xi$ is equal to two lattice spacings according to results from Ref.~\onlinecite{PowellMcKenzieReview:2011, Elstner:1993, Elstner:1994}. 
}
\label{Fig_diagram_DMFT_Triangle}
\end{center}
\end{figure}

Fig.~\ref{Fig_diagram_DMFT_Triangle} presents the $T-U$ phase diagram for DMFT and CDMFT Triangle. As explained at the end of Section~\ref{Sec_model_methods}, the DMFT results are close to those of CDMFT Triangle, as shown in Ref.~\onlinecite{Dang_Xu_Chen_Meng_Wessel_2015}, since the local nature of DMFT implicitly assumes that the lattice is magnetically frustrated so that long-wavelength magnetic correlations are unimportant. This manifests clearly in the locations of the Mott-transition end points, $(U_\mathrm{c},T_\mathrm{c}) \simeq (9.28,1/10)$ for DMFT and $(U_\mathrm{c},T_\mathrm{c}) \simeq (9.4,1/11)$ for CDMFT Triangle. This should be compared with DCA results from Ref.~\onlinecite{Dang_Xu_Chen_Meng_Wessel_2015} who found $(U_\mathrm{c},T_\mathrm{c}) \simeq (10.8,1/10)$ for single-site and $(U_\mathrm{c},T_\mathrm{c}) \simeq (8.2,1/10)$ for cluster sizes ranging from 3 to 6 sites. The supplemental material of Ref.~\onlinecite{Kokalj_McKenzie_triangle:_thermo:2013} contains an exhaustive table of the estimates obtained from various methods. Note also that the shape of the coexistence region for the Mott transition obtained from CDMFT, not shown here, is in agreement with experiments, as discussed in Ref.~\onlinecite{Liebsch:2009b}. 

Our focus here is the crossovers. These have been extensively studied for the triangular lattice, especially for the resistivity.\cite{Terletska:2011} In this context, quantum-critical scaling has been found theoretically,\cite{Dobrosavljevic:2013,Vucicevic_Tanaskovic:2015,Pustogow:2018,Eisenlohr_Lee_Vojta_2019} up to a factor of three in temperature above $(U_\mathrm{c},T_\mathrm{c})$, leading to detailed comparisons with experiments.\cite{FurukawaWidom:2015,Pustogow:2018,disorder_QCP_scaling_Mott_2019,Quantum_Critical_Scaling_Mott_2019,Critical_scaling_at_Mott_2019} The crossover lines in Fig.~\ref{Fig_diagram_DMFT_Triangle} have been defined differently from these last papers, as detailed in Section~\ref{Sec_Crossovers_characterization}. The Widom lines of Fig.~\ref{Fig_diagram_DMFT_Triangle} do not survive at high temperature, since it becomes impossible to find an inflexion point in $D$ or $\mathrm{Im}\,G^\mathrm{loc}(i\omega_0)$ as a function of $U$ following Eqs.~\eqref{Eq_Widom_D} and \eqref{Eq_Widom_G}. Their dependence on $U$ becomes linear instead of showing an inflexion point as in the top panels of Fig.~\ref{Fig_Widom_CDMFT_2x2_beta_11}. Whereas the Widom line disappears at temperatures between $0.2$ and $0.3$, the crossover associated with the opening of the Mott gap in the DMFT and CDMFT Triangle calculations can be identified over a larger range of temperatures and for all studied values of $U$ larger than $U_c$. 

The ability to experimentally observe these crossovers depends on the possible presence of an ordered phase at low temperature. In two dimensions, one may argue that long-range magnetic orders, \textit{e.g.} antiferromagnetism and spiral antiferromagnetism, are forbidden at finite temperature in systems with short-range interactions by the Mermin-Wagner theorem. However, the local mean magnetization $\langle \hat{S}_z^2 \rangle$ (with the spin operator $\hat{S}_z = \hat{n}_{i\uparrow} - \hat{n}_{i\downarrow}$) still affects various physical observables, such as the double occupancy $D$ that depends on $\langle \hat{S}_z^2 \rangle$ at half-filling as
\begin{equation}
\langle \hat{S}_z^2 \rangle = \langle (\hat{n}_{i\uparrow} - \hat{n}_{i\downarrow})^2 \rangle = 1-2D.
\label{Eq_magnetization_D}
\end{equation}
This can be proven using the Pauli exclusion principle $\hat{n}_{i\sigma}^2=\hat{n}_{i\sigma}$. More generally, the observability of the crossovers can be discussed by considering the value of the magnetic correlation length. High-temperature series expansion for the isotropic triangular lattice Heisenberg model find that the magnetic correlation length takes the value $\xi = 2$ lattice spacings at $T = 0.2\, J$, where $J$ is the Heisenberg antiferromagnetic exchange.\cite{Elstner:1993, Elstner:1994, PowellMcKenzieReview:2011} Using $J = 4t^2/U$ for a rough estimate of the value of the magnetic correlation length, we find $\xi=2$ at $T\simeq 1/11.76<T_\mathrm{c}$ for  $U=U_\mathrm{c}=9.4$, which means that magnetism is weak even at temperatures below the CDMT Triangle critical point. The dashed line in Fig~\ref{Fig_diagram_DMFT_Triangle} indicates where $\xi$ equals $2$ in the phase diagram. This is an overestimate of $\xi$ since for the values of $U$ of interest, we are not yet in the Heisenberg limit and ring exchange terms should be detrimental to the establishment of long-range order. The crossovers associated with the Widom line can thus be seen experimentally. They are in fact seen in organic materials.\cite{Powell:2005} In turn, the crossovers associated to the Mott line are also experimentally observable since they extend far above the low-temperature $\xi=2$ line. Our Mott line indeed fundamentally reflects the same physics as the metal-to-insulator crossover defined differently as a ``Quantum Widom line" and observed experimentally in Refs.~\onlinecite{FurukawaWidom:2015,Pustogow:2018,disorder_QCP_scaling_Mott_2019,Quantum_Critical_Scaling_Mott_2019,Critical_scaling_at_Mott_2019}.

\subsection{Square lattice}
\label{Sec_square_lattice}

\begin{figure}[h!]
\begin{center}
\includegraphics[width=0.48\textwidth]{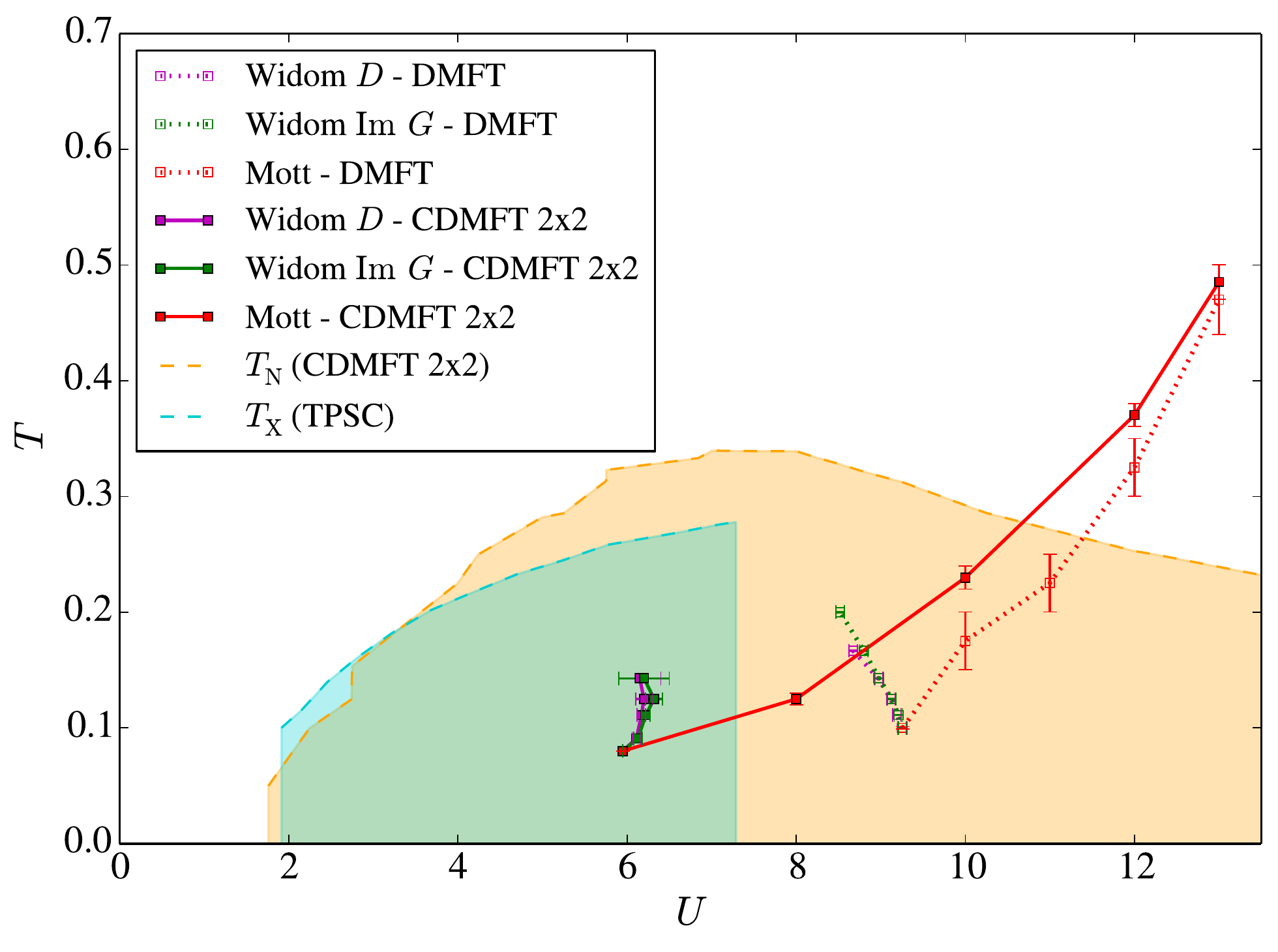}
\caption{$T-U$ phase diagram for DMFT Square and CDMFT $2\times 2$. 
``Widom $D$'' and ``Widom $\mathrm{Im}\,G$'', and ``Mott'' correspond to the Widom lines for the double occupancy and the imaginary part of the local Green's function taken at the first Matsubara frequency, and the Mott crossover, respectively. The TPSC crossover temperature data $T_\mathrm{X}$ are taken from Ref.~\onlinecite{Dare:2007} and the CDMFT $2\times 2$ N\'eel temperature data $T_\mathrm{N}$ from Ref.~\onlinecite{Fratino_Semon_Charlebois_Sordi_Tremblay_2017}. The domes are expected to extend from $U=0$ to $U\to +\infty$ but they are not drawn because of lack of data.}
\label{Fig_diagram_DCA_CDMFT}
\end{center}
\end{figure}

Compared to the isotropic triangular lattice, the non-frustrated square lattice is in the opposite limit, where long-wavelength antiferromagnetic fluctuations compete with the Mott transition, leading eventually to long-range order at zero temperature. It is then crucial to identify whether or not the Widom and Mott lines remain experimentally observable in this case. Fig.~\ref{Fig_diagram_DCA_CDMFT} presents the $T-U$ phase diagram for DMFT and CDMFT $2\times 2$. Even if the Mott transitions are not represented here, the location of the Mott critical points appears clearly method- and cluster-size- dependent. One has $(U_\mathrm{c},T_\mathrm{c}) \simeq (9.28,1/10)$ for DMFT and $(U_\mathrm{c},T_\mathrm{c}) \simeq (5.90\pm 0.05,0.06 \pm 0.005)$~\cite{Walsh_Semon_Poulin_Sordi_Tremblay_2019} for CDMFT $2\times 2$. These differences reflect the fact that the magnetic correlation length is large in the vicinity of the Mott critical point, so that different numerical techniques, treating non-local correlations differently, yield different results. In fact, the same high-temperature series expansion estimates mentioned for the triangular-lattice Heisenberg model previously give $\xi = 200$ lattice spacings at $T = 0.2\, J$ for the square-lattice Heisenberg model. 


Like the Mott critical points, the Widom lines of Fig.~\ref{Fig_diagram_DCA_CDMFT} depend on cluster size. Although one may think that they tend towards one another at high temperature, these crossovers do not survive at sufficiently high temperature (inflexion points cannot be found), to allow verification. Besides, it is the physics of antiferromagnetic fluctuations that becomes important, not crossovers associated to the Widom lines. To illustrate this, the Néel temperature, $T_\mathrm{N}$, for CDMFT $2\times 2$ and the crossover temperature $T_\mathrm{X}$ to the renormalized classical regime in the Two-Particle Self-Consistent (TPSC) approach~\cite{TPSC_book} are shown. The latter is defined as the temperature below which the magnetic correlation length $\xi$ starts increasing exponentially for a given value of $U$. At $T_\mathrm{X}$, and in the weak correlation regime, a pseudogap that is a precursor of the zero-temperature antiferromagnetic ground state appears.~\cite{Vilk:1995,Vilk:1997,Vilk:1999,Moukouri:2000}

Earlier work had found with several methods~\cite{Aichhorn:2014,Rohringer_Hafermann_Toschi_Katanin_Antipov_Katsnelson_Lichtenstein_Rubtsov_Held_2018} that the Mott critical point was below $T_\mathrm{N}$ and $T_\mathrm{X}$. Our contribution is to show that the Widom line is also below these temperatures. Even if there is no long-range antiferromagnetic order in two dimensions at finite temperature, $T_\mathrm{X}$ and $T_\mathrm{N}$ signal the effect of long-wavelength antiferromagnetic fluctuations that not only hide the Mott transition, but also the crossovers associated with the Widom line. Indeed, below these temperatures, quantities such as double occupancy behave qualitatively as if there was long-range order.~\cite{Fratino_Semon_Charlebois_Sordi_Tremblay_2017} While the DMFT line extends outside the region where antiferromagnetic fluctuations become important, this result is unphysical because including even short-range spatial correlations, as in CDMFT $2\times 2$, completely displaces the transition. Note also that the slope of the Widom line is positive for CDMFT $2\times 2$ on the square lattice while it is negative on the three-site cluster for the triangular lattice. This is understood from the fact that the Mott insulator in the former state has zero entropy while it has the entropy of a spin $1/2$ on the triangular cluster.~\cite{park:2008}  

It is important to notice, however, that the crossovers associated to the Mott line are certainly observable since, as for the triangular lattice in Fig.~\ref{Fig_diagram_DMFT_Triangle}, they extend to high temperatures where eventually the different methods give the same result.

\subsection{The high-temperature limit and the Mott line} 

\begin{figure}[h!]
\begin{center}
\includegraphics[width=0.48\textwidth]{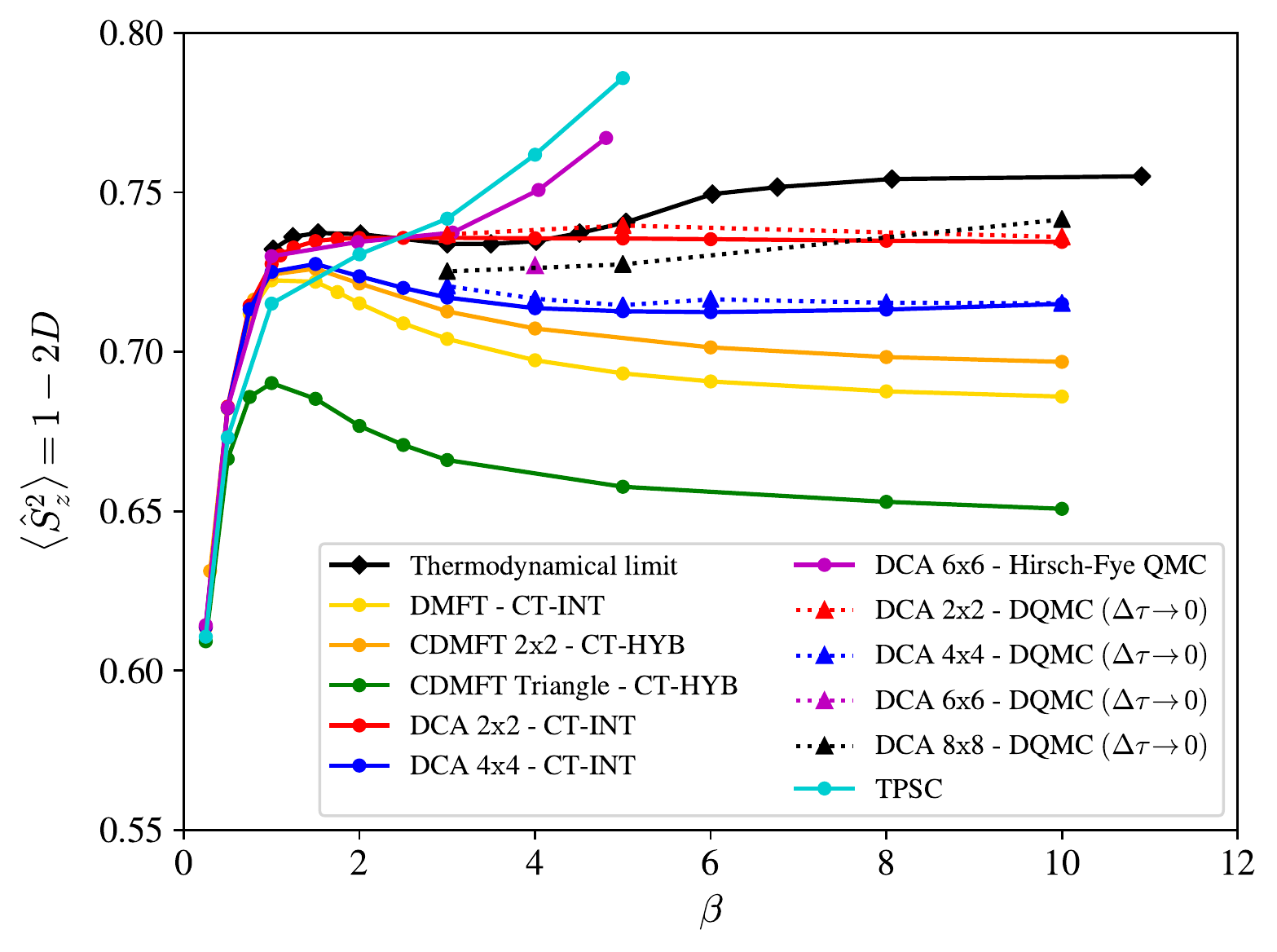}
\caption{Benchmarking of several numerical results as a function of temperature using the mean magnetization $\langle S_z^2 \rangle$ Eq.~\eqref{Eq_magnetization_D} as a function of $\beta$ at $U=4$. All theses methods agree with each other when $T \gtrsim t$ (or $\beta \lesssim 1$) because the thermal energy is large enough to mask the details of the lattice. The extrapolation to the thermodynamic limit is taken from Ref.~\onlinecite{Paiva:2001}, the DCA $6\times 6$ data are taken from Ref.~\onlinecite{Moukouri:2001} and the TPSC data are taken from Ref.~\onlinecite{Kyung:2003a}. The Determinant Quantum Monte Carlo (DQMC) data are extrapolated to the limit where the imaginary time discretization step $\Delta\tau$ goes to zero.}
\label{Fig_Benchmarking}
\end{center}
\end{figure}

At sufficiently high temperature, the results of all methods on all lattices should be identical since the physics becomes inherently local in that regime. Previous studies~\cite{Georges:2011} have shown that this occurs for $T \gtrsim t$. We benchmark this for $U=4$ in Fig.~\ref{Fig_Benchmarking}, where the mean local magnetization as a function of $\beta=1/T$ is found to be essentially the same for different cluster sizes, lattices and numerical methods when $T \gtrsim 1\equiv t$ (or $\beta \lesssim 1$). The extrapolation to the thermodynamic limit taken from Ref.~\onlinecite{Paiva:2001} exhibits the behavior expected from a buildup of antiferromagnetic correlations at low temperature and a subtle Pomeranchuk effect at intermediate to high temperature.\cite{Georges:1996}

\begin{figure}[h!]
\begin{center}
\includegraphics[width=0.48\textwidth]{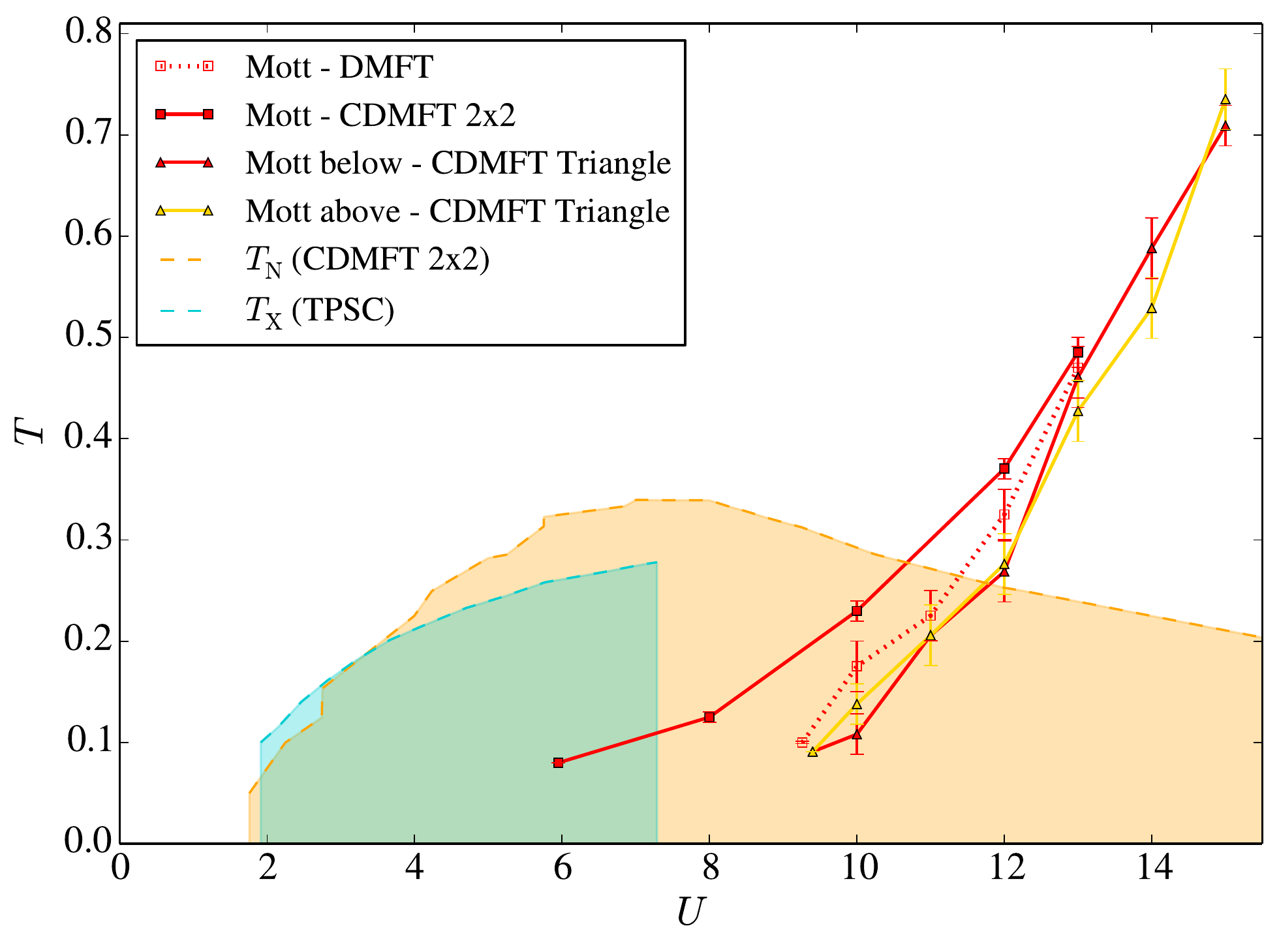}
\caption{$T-U$ phase diagram for the Mott lines on the triangular and the square lattice. They merge at high temperature. The region associated to antiferromagnetic fluctuations on the square lattice, described in Fig.~\ref{Fig_diagram_DCA_CDMFT}, is also shown for reference.}
\label{Fig_Crossovers}
\end{center}
\end{figure}

Since all previously encountered Mott lines survive at high temperature, they should also merge into one another at sufficiently high temperature. The question remains whether or not the required threshold temperature is as high as $T/t=1$. Fig.~\ref{Fig_Crossovers} shows that the Mott lines merge together for $T/t\gtrsim 0.45$.

\subsection{Analytic continuation and the crossover lines}

\begin{figure}[h!]
\begin{center}
\includegraphics[width=0.48\textwidth]{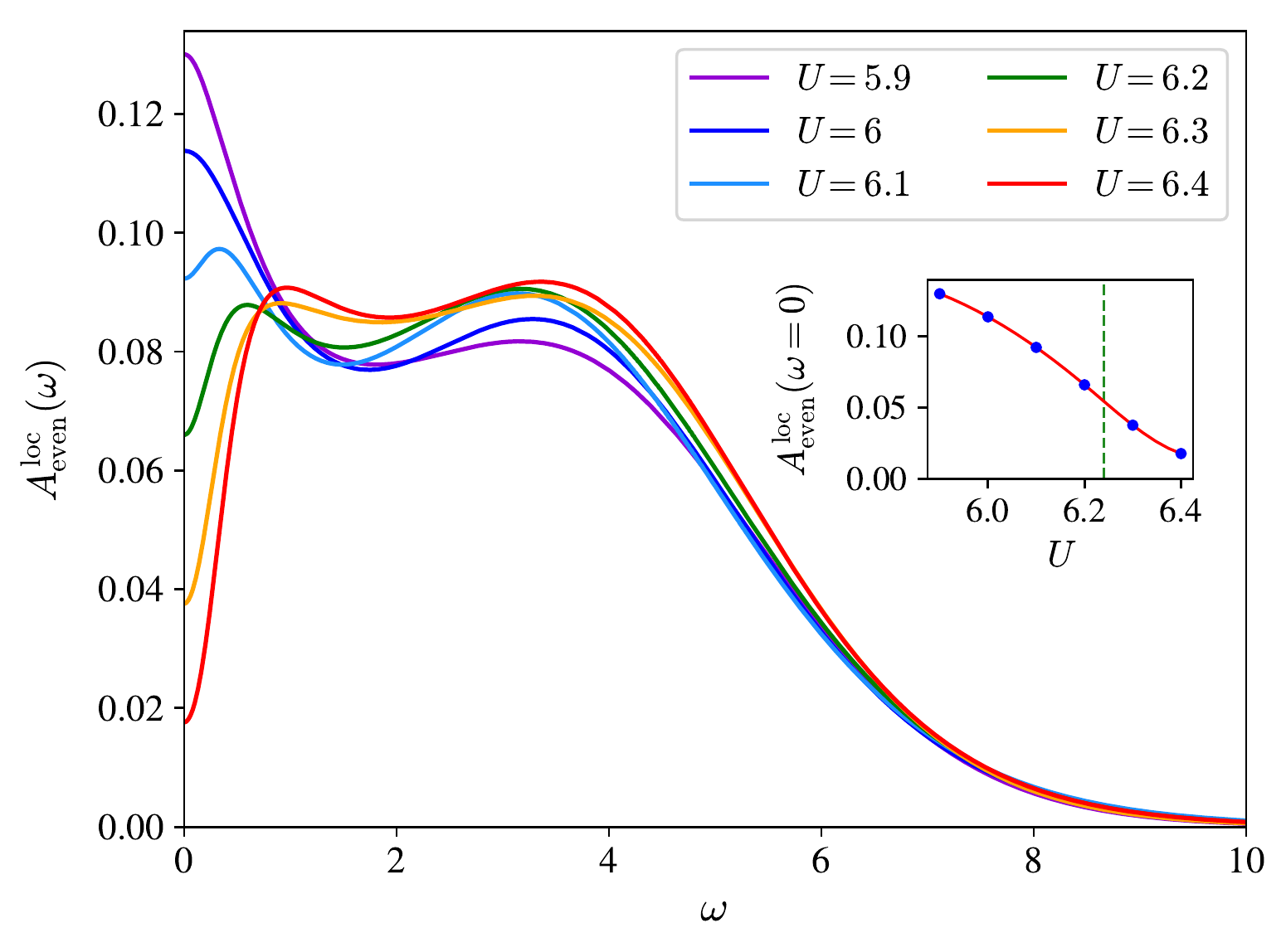}
\caption{Analytically continued local density of states of CDMFT $2\times 2$ on the square lattice at $\beta=9$ for different values of the interaction across the thermodynamic crossover in double occupancy ($U_\mathrm{W} \simeq 6.17$) and the dynamical crossover in the Green's function ($U_\mathrm{W}^\mathrm{d} \simeq 6.22$). The even parity of the particle-hole symmetric local density of states has been enforced by computing $A_\mathrm{even}^\mathrm{loc}(\omega) = (A^\mathrm{loc}(\omega)+A^\mathrm{loc}(-\omega))/2$. \textbf{Inset:} Value of the local density of states at $\omega=0$ as a function of $U$. An inflexion point (green dashed line) is observed at $U \simeq 6.24$, corresponding to another crossover that can be defined.}
\label{Fig_DOS_beta9}
\end{center}
\end{figure}

We use OmegaMaxEnt\cite{Bergeron:2015} to perform analytic continuation and gain a deeper understanding of the Widom and Mott lines. Fig.~\ref{Fig_DOS_beta9} presents the analytically continued local density of states of CDMFT $2\times 2$ at $\beta=9$ on the square lattice for different values of the interaction across the thermodynamic ($U_\mathrm{W} \simeq 6.17$) and dynamic ($U_\mathrm{W}^\mathrm{d} \simeq 6.22$) crossovers associated with the Widom line. The convexity of the low-energy density of states $A^\mathrm{loc}_\mathrm{even}(\omega)$ 
changes from concave (peak) to convex (dip) at $\omega=0$ across the Widom line, as previously observed at finite doping across the pseudogap temperature.\cite{Sordi:2012} This shows that there is a progressive transition towards a more insulating state as the interaction $U$ increases. The inset of Fig.~\ref{Fig_DOS_beta9} features another crossover characterized by an inflexion point in $A^\mathrm{loc}_\mathrm{even}(\omega=0)$ at $U\simeq 6.24$, which is of the same order of magnitude as $U_\mathrm{W}$ and $U_\mathrm{W}^\mathrm{d}$. Since this density of states is for intermediate interaction strength, there are shoulders that will develop into clear Hubbard bands at larger interaction strength.

\begin{figure}[h!]
\begin{center}
\includegraphics[width=0.48\textwidth]{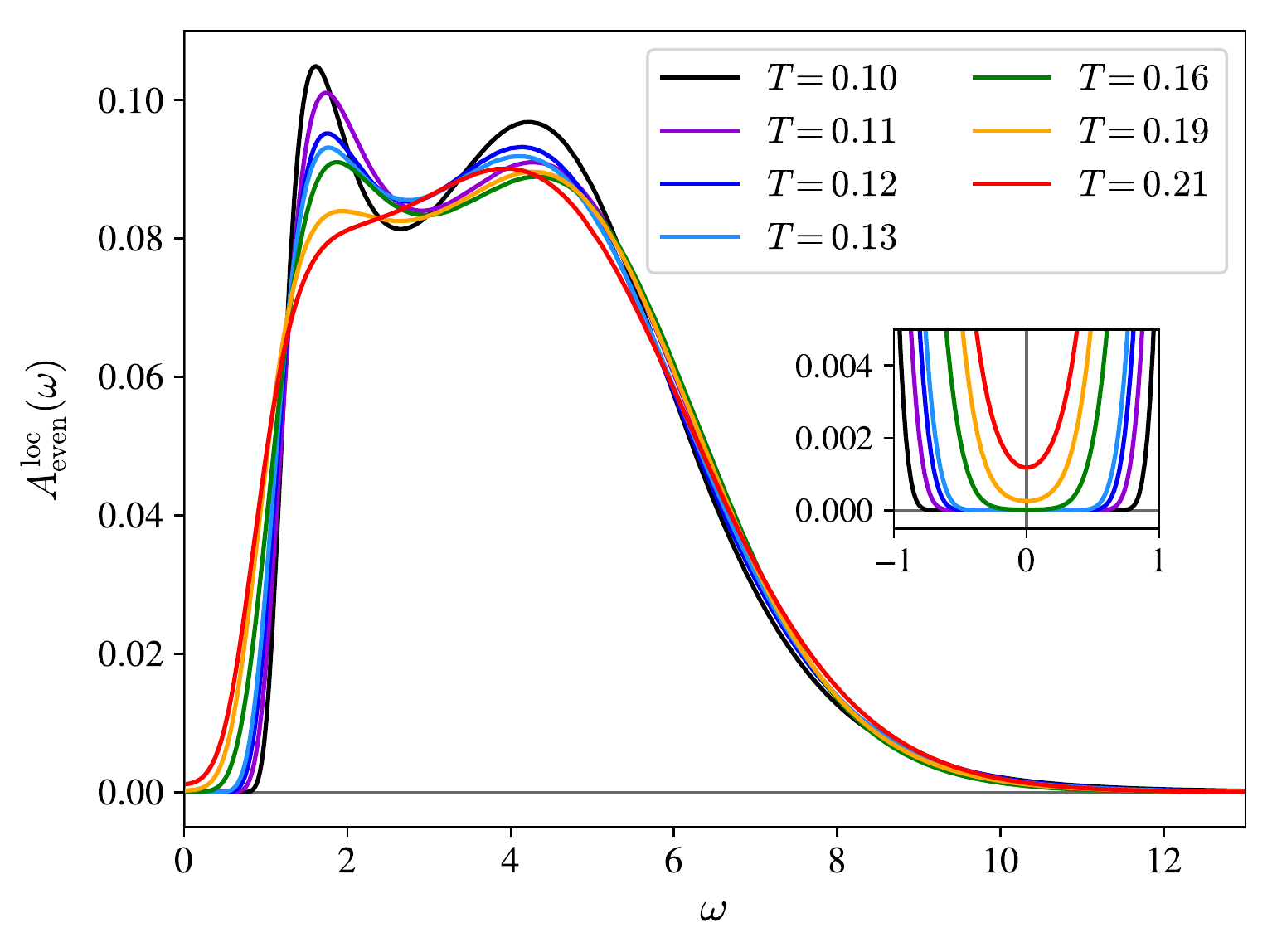}
\caption{Analytically continued local density of states for CDMFT $2\times 2$ on the square lattice at $U=8$ for different temperatures across the Mott line. The even parity of the particle-hole symmetric local density of states has been enforced, slightly correcting the analytic continuation. \textbf{Inset:} Closer look at the gap opening around $\omega=0$.}
\label{Fig_DOS_U8}
\end{center}
\end{figure}

To further understand the Mott line, consider Fig.~\ref{Fig_DOS_U8}. It displays the analytically continued local density of states of CDMFT $2\times 2$ at $U=8$ for different temperatures across the Mott line. The temperatures are sufficiently low to allow us to use analytic continuation. This is not possible for values of $U$ where the Mott line is at too high temperature. There the approach of Sec.~\ref{Sec:Mott_Line} is the only available one. The structure of the density of states in Fig.~\ref{Fig_DOS_U8} is more easily understood as it resembles the one found for the Hubbard model for large $U$:\cite{Moreo:1995, Preuss:1995, kyung:2006b} the first peak, whose width is of order $2J = 1$, comes from spin excitations whereas the second peak comes from Hubbard bands. The inset of Fig.~\ref{Fig_DOS_U8} shows that the gap is opening at $T\simeq 0.16$ while the Mott line that we found for $U=8$ with CDMFT $2\times 2$ was at $T\simeq 0.125 < 0.16$. The latter value coincides with the old results of Ref.~\onlinecite{Vekic:1993}. 
The fact that our Mott line is at lower temperature than that associated to the vanishing of the density of states at $\omega=0$ is expected, since our definition for the Mott line requires that the gap be opened enough so that $T \ll E_\mathrm{g}$ (see Eq.~\eqref{Eq_Mott_line_low_T_limit}). The Mott line found in Ref.~\onlinecite{Vekic:1993} with $6 \times 6$ DQMC calculations appears for a value of $U$ larger than about $6.5$, very close to that found with CDMFT $2\times 2$.

\section{Conclusion and summary}

The Mott transition controls the high-temperature normal-state crossovers that occur near the Widom line, where insulating behavior appears with increasing $U$, and near the Mott line, where a well-defined Mott gap is developed. The latter extends to very high temperature and values of $U$ larger than the critical $U_c$ for the Mott transition. It is a consequence of local physics.  On lattices that are highly frustrated, such as the triangular lattice, the first-order Mott transition extends to temperatures above the regime where spatial magnetic correlations set in.  Single-site DMFT gives a reasonable description of these crossovers. 

On the square lattice, which is often convenient to study theoretically because Quantum Monte Carlo methods have a smaller sign problem, the Widom line describes the same inflexion points and maxima in physical observables as those that occur on the triangular lattice, so they can serve as a convenient proxy for the results on frustrated lattices. The cluster-size dependence of the results however clearly shows that spatial correlations are important and that these crossovers are likely unobservable experimentally on non-frustrated lattices. The slope of the Widom line may also have a different sign on the square and on the triangular lattice for entropic reasons.~\cite{park:2008} 

On the square lattice, the first-order Mott transition is hidden by the onset of antiferromagnetic fluctuations and the associated pseudogap phenomena.~\cite{Aichhorn:2014,Rohringer_Hafermann_Toschi_Katanin_Antipov_Katsnelson_Lichtenstein_Rubtsov_Held_2018} Nevertheless, the Mott transition even in this case is not ``false''.~\cite{Aichhorn:2014} It always has a crucial role even in the latter case since its existence leads to the Mott line, that is always  observable at sufficiently high temperature above the strong antiferromagnetic fluctuation regime. The Mott transition on the square lattice appears in a normal state which is not stable at low temperature but which evolves continuously towards the correct normal state at high temperature, where the crossovers lie. This kind of behavior is common in other circumstances, for example when the high-temperature behavior of a metal is a Fermi liquid but its ground state is something else, such as a superconductor for example.

The Mott line will always be observable in principle at sufficiently high temperature and interaction strength, whatever the lattice. A different definition of the same crossover has been called a Quantum Widom line~\cite{Dobrosavljevic:2013,Vucicevic_Tanaskovic:2015,Eisenlohr_Lee_Vojta_2019} and has been observed in experiment,\cite{FurukawaWidom:2015,Pustogow:2018,disorder_QCP_scaling_Mott_2019,Quantum_Critical_Scaling_Mott_2019,Critical_scaling_at_Mott_2019} but we surmize that the scaling at the Quantum Widom line, which is restricted to a narrow temperature range, will not be observable if long-wavelength antiferromagnetic fluctuations extend too far above the Mott critical point.  

\begin{acknowledgments}
    We thank Vlad Dobrosavljevi\'c, Antoine Georges and Hanna Terletska for discussions. This work was partially supported by the Natural Sciences and Engineering Research Council (Canada) under grant RGPIN-2014-04584, the Fonds Nature et Technologie (Qu\'ebec), the Canada First Research Excellence Fund the Research Chair on the Theory of Quantum Materials (A.-M.S.T.), and the Canadian Institute for Advanced Research (A.-M.S.T.). Simulations were performed on computers provided by the Canadian Foundation for Innovation, the Minist\`ere de l’\'Education des Loisirs et du Sport (Qu\'ebec), Calcul Qu\'ebec, and Compute Canada.
\end{acknowledgments}


%

\end{document}